\documentclass[11pt,preprint]{aastex}
\usepackage{color}

\begin{document}

\title{UNCOVERING MECHANISMS OF CORONAL MAGNETISM VIA ADVANCED 3D MODELING OF FLARES AND ACTIVE REGIONS}

\author{Gregory Fleishman\altaffilmark{1}, Dale Gary\altaffilmark{1}, Gelu Nita\altaffilmark{1},
David Alexander\altaffilmark{2}, Markus Aschwanden\altaffilmark{3},
Tim Bastian\altaffilmark{4}, Hugh Hudson\altaffilmark{5}, Gordon
Hurford\altaffilmark{5}, Eduard Kontar\altaffilmark{6}, Dana
Longcope\altaffilmark{7}, Zoran Mikic\altaffilmark{8}, Marc
DeRosa\altaffilmark{3}, James Ryan\altaffilmark{9}, Stephen
White\altaffilmark{10}}

\affil{$^1$NJIT, $^2$Rice University, $^3$LMSAL, $^4$NRAO,
$^5$University of California, Berkeley, $^6$University of Glasgow,
$^7$Montana State University, $^8$PSI, $^9$University of New
Hampshire, $^{10}$AFRL/RVBXS}


\begin{abstract}
The coming decade will see the routine use of solar data of
unprecedented spatial and spectral resolution, time cadence, and
completeness.  To capitalize on the new  (or soon to be available)
facilities such as SDO, ATST and FASR, and the challenges they
present in the visualization and synthesis of multi-wavelength
datasets, we propose that realistic, sophisticated, 3D active region
and flare modeling is timely and critical, and will be a forefront
of coronal studies over the coming decade. To make such modeling a
reality, a broad, concerted effort is needed to capture the wealth
of information resulting from the data, develop a synergistic
modeling effort, and generate the necessary visualization,
interpretation and model-data comparison tools to accurately extract
the key physics.

\end{abstract}


\section{INTRODUCTION}

Solar activity, although energetically driven by subphotospheric
processes, depends critically on \emph{coronal magnetism}, which,
broadly speaking, includes magnetic field generation,
morphology/topology, evolution, and transformation into kinetic,
thermal, and nonthermal energies in the corona. While reliable
direct diagnostic information has been lacking, during the next
decade the situation must drastically change if we are to fully take
advantage of the large wealth of high resolution data and modeling
that is currently, or soon to be, available. New space- and
ground-based solar optical telescopes are already capable of precise
measurements of the photospheric magnetic field with sub-arcsecond
angular resolution and high temporal resolution. Being combined with
modern extrapolation algorithms, these data offer important clues on
the coronal magnetic field structure and evolution.

Owing to their finite angular resolution, sensitivity, observational
errors, and theoretical limitations, those extrapolations are not
unique, so the extrapolations require independent verification. 
An opportunity for quantitative verification through radio coronal
magnetography will be available during the coming decade when the
new generation of high-resolution solar-dedicated radio instruments,
including the expanded Owens Valley Solar Array (OVSA), upgraded
Siberian Solar Radio Telescope (SSRT), and Frequency Agile Solar
Radiotelescope (FASR), will become operational. There has also been
progress made in obtaining the coronal magnetic field of active
regions from advanced Stokes Polarimetry of infra red lines from the
HAO CoMP instrument \citep{Tomczyk_etal_2008}. Microwave radiation
from flares is produced by the gyrosynchrotron mechanism as
accelerated fast electrons gyrate in the coronal magnetic field. It
has recently been demonstrated using ideal (simulated) microwave
data, that the coronal magnetic field can, in principle, be reliably
recovered from the radio data on the flare dynamical timescale,
along with the key parameters of the thermal plasma and accelerated
electrons. The ability to detect the magnetic field and its changes
on the dynamic time scales is a critically needed element to uncover
fundamental physics driving solar flares, eruption, and activity.

\section{MODELING COMPONENTS}

Comprehensive modeling must include two closely related 
flows of effort: (i)  \emph{direct 3D modeling}
based on our theoretical knowledge and constrained by available
observations, which adopt some realistic physics and geometry and
predict/calculate observables,  and (ii)  \emph{robust diagnostics
tools}, which  achieve the complementary goal by starting from
observables in order to quantify the geometry and physical
processes.

Two major approaches  are available  to build the diagnostic tools:
\emph{true inversions}, which explicitly solve the inverse problem,
and \emph{forward fitting}, i.e., finding a number of free
parameters of a physically motivated model of the system from
fitting the model to observations.
These diagnostics tools, being developed, will provide us with an
array of the relevant coronal and flare physical parameters through
a detailed sophisticated analysis of the data: e.g., radio imaging
spectroscopy data for coronal plasma, accelerated electron, and
field conditions and X-ray data for flare dynamics and energetics.
We concentrate, here, on the solar flare modeling on dynamic time
scales (down to $\sim1$~s or so); as a byproduct, most of the model
components, described below, will also be applicable to the active
region modeling, see also
\verb"http://lws-trt.gsfc.nasa.gov/trt_sc20063dar.htm"

\vspace{0.22in}

\underline{\textbf{Direct Modeling.}}

Very few fully 3-dimensional models of solar flares have been
attempted up to the present time. Available models
\citep{Preka_Alis_1992, Kucera_etal_1993, Bastian_etal_1998,
Tzatzakis_etal_2008, Simoes_Costa_2006, Fl_etal_2009,
Simoes_Costa_2010} have used simplified, generic magnetic geometries
such as a dipole loop.  In order to make contact with observational
data, the next generation of models should employ realistic
geometries from extrapolations or long-time-evolution MHD
simulations
based on vector photospheric measurements of the magnetic and
velocity fields. 
As an example of complex concerted efforts leading eventually to the
relevant 3D modeling, we itemize below a number of required major
(but not all-inclusive) steps needed to develop realistic,
interactive, and adjustable 3D models of solar flares.

\vspace{0.22in}

\underline{Elements of direct flare modeling to be covered:}


\begin{enumerate}

%
%
%

    \item \textbf{A model of the pre-flare coronal plasma.}

The magnetic field, density, temperature and elemental abundance of
the pre-flare plasma is essential for modeling the subsequent solar
flare.  The basis of the magnetic field model would typically be a
non-linear force free extrapolation from a photospheric or
chromospheric vector magnetogram.  The expected situation is for a
low-beta hydrostatic equilibrium plasma to exist within this
magnetic structure.  The density and temperature structure would be
set by heating determined either by an accepted form
\citep{Schrijver_etal_2006, Lundquist_etal_2008} or through
comparison to EUV images. Alternatively a time-dependent,
slowly-evolving simulation could be used to supply the pre-flare
conditions.


    \item \textbf{The energy release and flaring site.}

According to current understanding, the rapid magnetic energy
release powering a solar flare results form fast magnetic
reconnection.  Where in the magnetic field this process will occur,
how much energy it will release and into which forms the energy will
be converted will be the subject of active research over the coming
decade.  The modeling proposed here will provide critical input and
constraints into these investigations.  One approach will be to
select flare-energized field lines or flux tubes based on the
observed locations of flare signatures, independent of  theoretical
considerations.  Comparing the results of subsequent modeling to
observations will then cast light on possible relations between the
magnetic environment and reconnection energy release. Alternatively,
a model for three-dimensional reconnection could be used to identify
the field lines onto which energy is released. Research in recent
years has suggested that non-fluid effects may play a critical role
in triggering reconnection \citep[see for example,][]{
Birn_Priest_2007}.  It might occur first where the width of the
current sheet has decreased to a scale comparable to the ion skin
depth, permitting kinetic effects to become effective.

The largest flares tend to be associated with dynamical eruptions
called coronal mass ejections (CMEs).  The exact relationship
between the flare and the CME is still the subject of ongoing
research, but it is generally agreed that the magnetic configuration
is far from equilibrium at the time of the flare.

 \item \textbf{Dynamics of the thermal plasma in the flaring loop}
 (or in the flaring loop system), i.e., prescribing the evolving electron number density,
 elemental/ion composition, and temperature to each voxel.

The density and temperature may evolve in time as specified by
hydrodynamic response to the flare energy release; ideally the loop
will be embedded in a global (pre-flare) coronal model (1). At the
level of more advanced modeling the inhomogeneous active region
atmosphere is specified self-consistently with the magnetic
structure, heating sources (including those driven by the flare),
and cooling \citep[see, e.g.,][]{Mok_etal_2008}. In this case (in
place of populating the voxels) the magnetic field model is
self-consistently coupled with the thermal plasma distribution, so
the data cubes describing the thermal plasma distribution must be
imported into the flare modeling tool along with the magnetic field
data cube.

 \item \textbf{Populating the loop by evolving fast accelerated electrons}, which
eventually must be determined from the time-dependent solutions of
the transport equation (for flare modeling; not needed in case of
active regions).

The full problem is far from its final solution. The problem
includes two major ingredients: particle acceleration and particle
transport. 
Currently, there is no consensus about the main acceleration
mechanisms operating in flares, although there are numerous models
capable of successfully accounting for some of the observed
properties of the accelerated particles. The modeling tools must be
capable of accommodating arbitrary outcomes of (either analytical or
numerical) external models describing the particle acceleration.
Once the accelerator has been set up, the consequent transport
mechanism is basically known: it is defined by the magnetic field
line structure and thermal plasma, which is known within a given
model. An additional but typically unknown important ingredient,
which can affect the transport, is  wave turbulence capable of
angular scattering the particles.  The presence and amplitude of
such turbulence would follow either from the energy release model
(2) or be set empirically using other observational
input, such as spectral line widths.

 \item   \textbf{Calculation of the thermal and nonthermal} (HXR,
gamma-ray, radio etc.) \textbf{emission characteristics} in each
voxel and solution of the radiation transfer equations along all
selected lines of sight, thus forming  data cubes of the emission
for the preselected viewing angle.

These tasks, although based on the well understood and well
developed theory, are frequently computationally demanding. So
specific efforts for minimizing the computation time and fully
optimized computing codes are critically needed.

\vspace{1cm}
 \item \textbf{Developing powerful user-friendly visualization tools for the
variety of model-derived 3D data cubes.}

The visualization tools are needed to fully understand the structure
and properties of the 3D objects under study. In addition, similar
tools are needed to look at the imaging spectroscopy data as the
relevant data volumes are often larger than those available from
individual context instruments, and require spherical geometry, so
that advanced tools are needed for data coalignment, viewing and
analysis.

 \item   \textbf{Folding the data cubes through the instrumental response functions for direct
comparison with observations.}

\end{enumerate}

Most of the items mentioned represent major sub-projects within the
program of the overall 3D modeling effort.  The larger effort
requires coordination of model standards and compatibility of the
model data formats and consistently defined physical assumptions
between the component parts, to enable these sub-projects to mesh
seamlessly. Ultimately, these simulated models will be analyzed by
the forward fitting or inversion tools, early versions of which are
already in development (e.g., Fleishman et al. 2009). They are
intended for reliably deriving the physical parameters of the
emission region. This direct modeling, coupled with the forward
fitting or true inversions, is a critically important step for
validating the diagnostics tools prior to their application to real
observational data.

\vspace{0.22in}

\underline{\textbf{Forward Fitting Diagnostics.}}

Robust diagnostics, understood as the determination of physical
parameters of a system under study from arrays of observed
parameters, is a key outstanding problem in Solar Physics. In some
cases, e.g., in the hard X-ray (HXR) range, regularized true
inversions can work well \citep[e.g.,][]{Kontar_etal_2004}. In most
cases, however, true inversions fail because of the highly nonlinear
nature of the physical systems they are trying to extract
information from. In such cases,  forward fitting can often be
successfully used in place of true inversions. Although we
specifically discuss below the forward fitting of the imaging
spectroscopy of the microwave data, most of the discussion applies
to imaging spectroscopy data in other wavelengths as well.

Assume that we have a sequence of spatially resolved spectra (both
intensity and polarization data) from a solar flare (e.g., one
spectrum per pixel). Then, we can fit the data to a model spectrum
pixel by pixel to derive physical parameters of the source.

\vspace{0.22in}

\underline{This forward fitting includes the following elements.}

\begin{enumerate}
    \item  \textbf{Identify the model source function based on the radiation
    mechanism involved.}

In the case of microwave emission from solar flares the emission
mechanism is generally known: it is gyrosynchrotron (GS) emission
with a free-free contribution in some cases
\citep[e.g.,][]{Bastian_etal_1998}. Although the corresponding
emission and absorption coefficients are known theoretically, the
exact GS formulae are very computationally expensive and cannot be
used in practice as the forward fitting input. Fortunately, much
faster codes giving the same accuracy have recently been developed
by \cite{Fl_Kuzn_2010}. These codes, being fast, precise, and
applicable for a broad range of regimes including anisotropic
distributions of fast electrons imply a breakthrough in both 3D
direct modeling and the forward fitting, allowing forward fittings
of large bodies of data over reasonable time.

    \item \textbf{Identify the fitting procedure resulting in fast and reliable
    finding of the true source parameters.}

The problem is that most of the minimization algorithms  often find
a \emph{local} minimum of the normalized residual (or of the reduced
chi-square), while the ultimate goal of the fitting  is to identify
the \emph{global} minimum. So far, we have  determined that the
simplex algorithm is very efficient in finding a local minimum.
Then, it needs to be 'shaken' for the simplex solution to overcome
any local minima and continue downhill towards the global minimum (a
version of the stimulated annealing approach). Further efforts in
optimizing the minimization algorithm are still needed especially
for more complex cases when the number of the free model parameters
is large.

    \item \textbf{After-fitting inspection of the results.}

Even when the algorithm performance is very good overall, there is a
non-zero probability that the algorithm fails to find the true
solution in some pixels. The post-processing must be able to
identify and flag/remove those pixels.

    \item \textbf{Interactive methods (similar to those used for the direct modeling)
    to deduce changes to model parameters based on observed mismatch.}

\end{enumerate}

This sequential forward fitting must pave the way towards a global
fitting in which a global source model is fitted to the whole body
of the observational data (a multidimensional data cube containing
spectra, light curves, and evolving spatial structure). Ultimately,
the above procedures are iterated by quantitative means (to be
determined), to adjust the model to match observations. Note that
the model is adjusted to simultaneously match all available
(multi-wavelength) observations.

\section{THEORETICAL INPUT}

The outlined modeling efforts require further development of the
theory;  the following key input is particularly important.

\begin{enumerate}
    \item \textbf{Magnetic field extrapolations, magnetic reconnection, and energy
release.}


The present state of the art, non-linear force-free extrapolation
from photospheric vector data, is known to suffer from inadequacies
which must be overcome, or at least addressed, if we are to make
progress in the coming decade.  The photospheric field does not
itself satisfy the force-free conditions used to extrapolate it.  It
must either be somehow modified or replaced entirely with vector
field measurements from higher in the atmosphere
\citep{Metcalf_etal_1995, DeRosa_etal_2009}.   There are numerous
other sources of information into the coronal magnetic field, such
as EUV images \citep[see][]{Malanushenko_etal_2009}, radio emission,
coronal polarimetry \citep{Lin_etal_2004, Tomczyk_etal_2008} or loop
oscillations \citep{Aschwanden_etal_1999, Nakariakov_etal_1999}.
These data are either indirect (EUV loops or oscillations) or in a
form (sparse or integrated) not easily incorporated into the
traditional formalism of extrapolation.  Nevertheless, they
constitute valuable data on the coronal field and should not be
discarded entirely.  Innovative techniques must be developed for
incorporating such data into coronal field models.  The modeling
effort here proposed will, in the end, provide constraints on the
structure of the coronal magnetic field and it would be valuable to
be able to incorporate this back into the magnetic model itself.

The very nature of upward extrapolation is poorly suited to
identifying and resolving thin magnetic structure far from the
boundary.  Such structures, current sheets in particular, are
believed to play an essential role in the rapid energy release
powering flares.  In order to make genuine progress in understanding
flares new coronal models must be developed which are capable of
including these structures and their associated free energy.

    \item   \textbf{Turbulence generation, evolution, and parametrization
of wave-particle interactions.}

Turbulence is a highly important (but elusive for direct probing)
element of the coronal plasma. It can play a key role in particle
acceleration and transport and in the generation of electromagnetic
emission. Since it often cannot be measured, the role of the theory
here is crucial to provide the necessary relevant input for the
modeling.

    \item  \textbf{Particle acceleration and transport including simplified fast solutions.}

Particle acceleration remains an outstanding problem in solar
flares. Any progress in understanding the acceleration mechanisms
and how they work in realistic geometries will be exceedingly
valuable for the modeling discussed above. The accelerator plays a
role of the source function for the particle transport. Although the
transport equations are generally known, their numerical solution is
computationally demanding; thus, like in case with radiation
coefficients, simplified and optimized (but still accurate)
solutions are needed.

    \item  \textbf{Radiation processes including emissivity, absorption coefficients,
    and radiation transfer solutions; including fast codes.}

Although many radiative processes are well studied, there is a need
to optimize the computing codes for speed in many cases
\citep[e.g.,][]{Fl_Kuzn_2010}. In addition, new emission processes
have recently been identified, e.g., diffusive synchrotron radiation
\citep{li_fleishman_2009}, which can give dominant contribution in
the very site of the stochastic acceleration. Thus, both new regimes
and mechanisms of emission and efficient computing codes must be
developed to provide valuable input to the modeling.

\end{enumerate}

\section{EXPECTED RESULTS AND OUTSTANDING QUESTIONS}


The modeling effort outlined  above would have a broad impact on the
field of solar astronomy.  Its most immediate results would be in
progress toward answering the following questions.

\begin{itemize}

    \item  Determination of the location of the energy release,
    the means by which magnetic energy is rapidly released, magnetic field reconfigurations, and
 the mechanisms of particle acceleration.

    \item  Quantitative verification of magnetic field
extrapolations, and refinement of extrapolation methods.

    \item Quantitative understanding of the hydrodynamic response of the atmosphere to
energy input.

    \item   Quantitative understanding of accelerated particle
distributions, including energy, pitch angle, and relative
proportions of thermal/nonthermal partition.

    \item   The role of turbulence and wave-particle interactions on transport of
particles and energy.

\end{itemize}

\newpage


\section{BROADER IMPLICATIONS}

The  modeling efforts we have outlined can only bring fundamental
knowledge about flare/active region physics if used in conjunction
with high-resolution modern observations. Key observations of the
coronal plasma parameters can only be made by radio instruments that
combine high sensitivity, temporal, spatial, and spectral
resolution, which are unavailable now. A small part of the required
science will be possible soon with the upgraded OVSA instrument (the
funding of the upgrade just [10/01/2010] started; anticipated start
of the upgraded instrument operation is fall, 2013). However, the
full required capability has to await the completion of the full
FASR, see Concept Papers submitted by Bastian et al., Gary et al.,
White et al.

\section{CONCLUSIONS}

As is clear from this Concept Paper, sophisticated development of
three elements will be needed over the coming decade to
significantly improve our understanding of the coronal magnetism,
turbulence generation, and particle acceleration: (i) direct 3D
modeling; (ii) theory; and (iii) iterative methods and tools for
analysis of new, high-resolution, observations through forward
fitting.

\newpage

\bibliographystyle{apj} 
\bibliography{WP_bib,fleishman}

\begin{thebibliography}{21}
\expandafter\ifx\csname natexlab\endcsname\relax\def\natexlab#1{#1}\fi

\bibitem[{{Aschwanden} {et~al.}(1999){Aschwanden}, {Fletcher}, {Schrijver}, \&
  {Alexander}}]{Aschwanden_etal_1999}
{Aschwanden}, M.~J., {Fletcher}, L., {Schrijver}, C.~J., \& {Alexander}, D.
  1999, \apj, 520, 880

\bibitem[{{Bastian} {et~al.}(1998){Bastian}, {Benz}, \&
  {Gary}}]{Bastian_etal_1998}
{Bastian}, T.~S., {Benz}, A.~O., \& {Gary}, D.~E. 1998, \araa, 36, 131

\bibitem[{{Birn} \& {Priest}(2007)}]{Birn_Priest_2007}
{Birn}, J. \& {Priest}, E.~R. 2007, {Reconnection of magnetic fields :
  magnetohydrodynamics and collisionless theory and observations}, ed. {Birn,
  J.~\& Priest, E.~R.}

\bibitem[{{De Rosa} {et~al.}(2009){De Rosa}, {Schrijver}, {Barnes}, {Leka},
  {Lites}, {Aschwanden}, {Amari}, {Canou}, {McTiernan}, {R{\'e}gnier},
  {Thalmann}, {Valori}, {Wheatland}, {Wiegelmann}, {Cheung}, {Conlon},
  {Fuhrmann}, {Inhester}, \& {Tadesse}}]{DeRosa_etal_2009}
{De Rosa}, M.~L., {Schrijver}, C.~J., {Barnes}, G., {Leka}, K.~D., {Lites},
  B.~W., {Aschwanden}, M.~J., {Amari}, T., {Canou}, A., {McTiernan}, J.~M.,
  {R{\'e}gnier}, S., {Thalmann}, J.~K., {Valori}, G., {Wheatland}, M.~S.,
  {Wiegelmann}, T., {Cheung}, M.~C.~M., {Conlon}, P.~A., {Fuhrmann}, M.,
  {Inhester}, B., \& {Tadesse}, T. 2009, \apj, 696, 1780

\bibitem[{{Fleishman} \& {Kuznetsov}(2010)}]{Fl_Kuzn_2010}
{Fleishman}, G.~D. \& {Kuznetsov}, A.~A. 2010, \apj, 721, 1127

\bibitem[{{Fleishman} {et~al.}(2009){Fleishman}, {Nita}, \&
  {Gary}}]{Fl_etal_2009}
{Fleishman}, G.~D., {Nita}, G.~M., \& {Gary}, D.~E. 2009, \apjl, 698, L183

\bibitem[{{Kontar} {et~al.}(2004){Kontar}, {Piana}, {Massone}, {Emslie}, \&
  {Brown}}]{Kontar_etal_2004}
{Kontar}, E.~P., {Piana}, M., {Massone}, A.~M., {Emslie}, A.~G., \& {Brown},
  J.~C. 2004, \solphys, 225, 293

\bibitem[{{Kucera} {et~al.}(1993){Kucera}, {Dulk}, {Kiplinger}, {Winglee},
  {Bastian}, \& {Graeter}}]{Kucera_etal_1993}
{Kucera}, T.~A., {Dulk}, G.~A., {Kiplinger}, A.~L., {Winglee}, R.~M.,
  {Bastian}, T.~S., \& {Graeter}, M. 1993, \apj, 412, 853

\bibitem[{{Li} \& {Fleishman}(2009)}]{li_fleishman_2009}
{Li}, Y. \& {Fleishman}, G.~D. 2009, \apjl, 701, L52

\bibitem[{{Lin} {et~al.}(2004){Lin}, {Kuhn}, \& {Coulter}}]{Lin_etal_2004}
{Lin}, H., {Kuhn}, J.~R., \& {Coulter}, R. 2004, \apjl, 613, L177

\bibitem[{{Lundquist} {et~al.}(2008){Lundquist}, {Fisher}, {Metcalf}, {Leka},
  \& {McTiernan}}]{Lundquist_etal_2008}
{Lundquist}, L.~L., {Fisher}, G.~H., {Metcalf}, T.~R., {Leka}, K.~D., \&
  {McTiernan}, J.~M. 2008, \apj, 689, 1388

\bibitem[{{Malanushenko} {et~al.}(2009){Malanushenko}, {Longcope}, \&
  {McKenzie}}]{Malanushenko_etal_2009}
{Malanushenko}, A., {Longcope}, D.~W., \& {McKenzie}, D.~E. 2009, \apj, 707,
  1044

\bibitem[{Metcalf {et~al.}(1995)Metcalf, Jiao, Mc{C}lymont, Canfield, \&
  Uitenbroek}]{Metcalf_etal_1995}
Metcalf, T.~R., Jiao, L., Mc{C}lymont, A.~N., Canfield, R.~C., \& Uitenbroek,
  H. 1995, 439, 474

\bibitem[{{Mok} {et~al.}(2008){Mok}, {Miki{\'c}}, {Lionello}, \&
  {Linker}}]{Mok_etal_2008}
{Mok}, Y., {Miki{\'c}}, Z., {Lionello}, R., \& {Linker}, J.~A. 2008, \apjl,
  679, L161

\bibitem[{{Nakariakov} {et~al.}(1999){Nakariakov}, {Ofman}, {Deluca},
  {Roberts}, \& {Davila}}]{Nakariakov_etal_1999}
{Nakariakov}, V.~M., {Ofman}, L., {Deluca}, E.~E., {Roberts}, B., \& {Davila},
  J.~M. 1999, Science, 285, 862

\bibitem[{{Preka-Papadema} \& {Alissandrakis}(1992)}]{Preka_Alis_1992}
{Preka-Papadema}, P. \& {Alissandrakis}, C.~E. 1992, \aap, 257, 307

\bibitem[{{Schrijver} {et~al.}(2006){Schrijver}, {De Rosa}, {Metcalf}, {Liu},
  {McTiernan}, {R{\'e}gnier}, {Valori}, {Wheatland}, \&
  {Wiegelmann}}]{Schrijver_etal_2006}
{Schrijver}, C.~J., {De Rosa}, M.~L., {Metcalf}, T.~R., {Liu}, Y., {McTiernan},
  J., {R{\'e}gnier}, S., {Valori}, G., {Wheatland}, M.~S., \& {Wiegelmann}, T.
  2006, \solphys, 235, 161

\bibitem[{{Sim{\~o}es} \& {Costa}(2006)}]{Simoes_Costa_2006}
{Sim{\~o}es}, P.~J.~A. \& {Costa}, J.~E.~R. 2006, \aap, 453, 729

\bibitem[{{Sim{\~o}es} \& {Costa}(2010)}]{Simoes_Costa_2010}
---. 2010, \solphys, 266, 109

\bibitem[{{Tomczyk} {et~al.}(2008){Tomczyk}, {Card}, {Darnell}, {Elmore},
  {Lull}, {Nelson}, {Streander}, {Burkepile}, {Casini}, \&
  {Judge}}]{Tomczyk_etal_2008}
{Tomczyk}, S., {Card}, G.~L., {Darnell}, T., {Elmore}, D.~F., {Lull}, R.,
  {Nelson}, P.~G., {Streander}, K.~V., {Burkepile}, J., {Casini}, R., \&
  {Judge}, P.~G. 2008, \solphys, 247, 411

\bibitem[{{Tzatzakis} {et~al.}(2008){Tzatzakis}, {Nindos}, \&
  {Alissandrakis}}]{Tzatzakis_etal_2008}
{Tzatzakis}, V., {Nindos}, A., \& {Alissandrakis}, C.~E. 2008, \solphys, 253,
  79

\end{thebibliography}

\end{document}